# Relativistic Killingbeck Energy States Under an External Magnetic Fields

M. Eshghi[1,*], H. Mehraban[2], S.M. Ikhdair[3,4]

[1] *Researchers and Elite Club, Central Tehran Branch, Islamic Azad University, Tehran, Iran*

[2] *Faculty of Physics, Semnan University, Semnan, Iran*

[3] *Department of Physics, Faculty of Science, An-Najah National University, Nablus, West Bank, Palestine*

[4] *Department of Electrical Engineering, Near East University, Nicosia, Northern Cyprus, Mersin 10, Turkey*

**Abstract**

We address the behavior of the Dirac equation with the Killingbeck radial potential including the external magnetic and Aharonov-Bohm (AB) flux fields. The spin and pseudo-spin symmetries are considered. The correct bound state spectra and their corresponding wave functions are obtained. We seek such a solution using the biconfluent Heun's differential equation method. Further, we give some of our results at the end of this study. Our final results can be reduced to their non-relativistic forms by simply using some appropriate transformations. The spectra, in the spin and pseudo-spin symmetries, are very similar with a slight difference in energy spacing between different states.

## 1. Introduction

The solution of the relativistic equation for massive fermions is still a very challenging problem even after it has been derived more than 80 years ago and utilized. In fact, this wave equation has been received a rapidly growing importance in so many physical sciences. For example, it is used to describe the behavior of nucleons in nuclei when studying materials such as graphene [1,2], heavy ions spectroscopy and more recently in laser–matter interaction (see Ref. [3] and the references therein), optical lattices [4-6] and 2D and 3D topological insulators [7,8].

---

[*] **Corresponding** Author Email: *eshgi54@gmail.com*; *m.eshghi@semnan.ac.ir*

On the other hand, symmetry plays a fundamental role in physics. The idea about spin and pseudo-spin (p-spin) symmetries with the nuclear shell model has been introduced five decades ago [9, 10], and has been widely used in explaining a number of phenomena in nuclear physics and its related areas. Namely, certain aspects of deformed and exotic nuclei have been investigated by means of the concepts of spin and p-spin symmetries [9, 10].

In 1975, Bell and Ruegg [11] showed that p-spin symmetry is a relativistic symmetry of the Dirac Hamiltonian that occurs when $V(r) \approx -S(r)$ in magnitude of the repulsive time-like vector potential $V(r)$ and an attractive scalar potential $S(r)$ with opposite signs [12-14]. This case is called the p-spin symmetry.

At the beginning, the p-spin symmetry was considered in the nonrelativistic framework [10]. However, in the 1990s, Blokhin *et al* found a connection between p-spin symmetry and relativistic mean field theory [15-17]. In 1997, the relativistic feature of the symmetry was recognized and revived by Ginocchio [12]. With this mention of the relativistic feature of the p-spin symmetry, he also showed that if the scalar potential is equal to the time-like vector potential, $S(r) \approx V(r)$, then the Dirac Hamiltonian has also the so-called spin symmetry [18] where, this symmetry was applied to explain the suppression of spin-orbit splitting in states of mesons (with a heavy and a light quark) and antinucleons [18]. This symmetry also led to understanding of magic numbers in nuclei [19]. Further, the p-spin symmetry was introduced to explain the near degeneracy of some single-particle levels near the Fermi surface, in exotic nuclei [20], and to establish an affective nuclear shell-model scheme [21].

In fact, the p-spin doublets were suggested that based on the small energy difference between pairs of nuclear single-particle states, a quasi degeneracy of single nucleon doublets, with nonrelativistic quantum numbers $(n_r, \ell, j = \ell + 1/2)$ and $(n_r - 1, \ell + 2, j = \ell + 3/2)$, where $n_r, \ell$ and $j$ are the single-nucleon radial, orbital and total angular momentum quantum numbers, respectively [9, 10]. The total angular momentum $j = \tilde{\ell} + \tilde{s}$ with $\tilde{\ell} = \ell + 1$ is a pseudo-angular momentum and $\tilde{s}$ is p-spin angular momentum [22-26].

These degenerative single-nucleon levels are considered as a doublet structure with $\left(\tilde{n}=n-1, \tilde{\ell}=\ell+1, j=\tilde{\ell}\pm 1/2\right)$ where $\tilde{\ell}$ and $\tilde{s}=1/2$ are pseudo-orbital angular momentum and p-spin quantum numbers, respectively. In this regard, the deformation, identical bands, super-deformation and magnetic moment in the nuclear structures have been successfully explained by using this doublets structure [27-30].

One notable feature of these symmetries is the suppression of either the spin-orbit or the so-called p-spin-orbit couplings that are shown in the second-order equations for the upper/lower Dirac spinor components, respectively. In other works, the nature of the spin and p-spin symmetries were considered in the framework of perturbation theory [31] and the energy splittings of the p-spin doublets can be investigated as a result of small perturbation around the Hamiltonian with the potential [31] (see Refs. [32, 33].

Now, after the pioneering work of Ginocchio, the p-spin and spin symmetries have been investigated by solving the relativistic Dirac equation for a spin $1/2$ particle by means of different methods for exactly solvable potentials. Alhaidari *et al* [34] have showed in detail physical interpretation on the 3D Dirac equation in the case of spin symmetric limit $V(r)-S(r)=0$ and p-spin symmetric limit $V(r)+S(r)=0$. Meng *et al* [35] have investigated that the p-spin symmetry is exact under the general condition $d\left(S(r)+V(r)\right)/dr=0$ where it can be approximately satisfied in exotic nuclei with highly diffused potentials. Based on this limit, the p-spin SU(2) algebra was established [36] and, with the same origin, the spin symmetry in single anti-nucleon spectrum was proposed and investigated [37, 38]. In fact, the spin and the p-spin symmetries are SU(2) symmetries of a Dirac Hamiltonian with time-like vector and scalar potentials realized when the difference between the potentials or their sum is a constant [11]. In addition, for the two symmetries, the Dirac Hamiltonian is invariant under the SU(2) algebra [11].

Therefore, if $d\Sigma(r)/dr$ is equal to zero, where $\Sigma(r)=S(r)+V(r)=constant$, we have the p-spin symmetry while if $d\Delta(r)/dr$ is equal to zero, where $\Delta(r)=V(r)-S(r)=constant$, we have the spin symmetry [35, 37, 39]. In fact, when the potentials are spherical, the Dirac equation is said to have the spin or p-spin symmetry corresponding to the same or opposite sign.

For a further review on this subject, the reader can refer to the recent works by Shen *et al* [40], Alberto *et al* [41], Lisboa *et al* [42], Marcos *et al* [43] and other works [44-47].

Also, in recent years, the exact analytical solutions of the Dirac equation have been extensively obtained, by using various methods in the presence of the spin and p-spin symmetries (see, for example, [48-58]). On the other hand, some authors have studied both the non-relativistic and relativistic bound states with some potential models such as Cornell [59], harmonic oscillator [60], sum of harmonic oscillator and Cornell [61], the superposition of pseudo-harmonic, linear and Columbic potential forms [62], the Coulomb plus linear [63], the anharmonic oscillator [64], the Hulthen [65], the Coulomb-like [66] and the Killingbeck potentials [67] in the presence of the external magnetic and AB flux fields.

Recently, Neyazi *et al* [68] have investigated the triaxial nuclei with the Killingbeck potential model.

In this work, we attempt to solve the Dirac equation under the spin and p-spin symmetries with the Killingbeck potential model [69] which is of great importance, particularly, in particle physics. It has the general form:

$$V(r) = ar^2 + br - \frac{c}{r}, \qquad (1)$$

which is mainly used to study the splitting of the relativistic energy eigenvalues in the presence of the external magnetic and *AB* flux fields.

The organization of the present paper is as follows. In Section 2, we solve the (2+1) dimensional Dirac equation with potential model (1) in the presence of external magnetic and *AB* flux fields using two different methods. We study both the spin and pseudo-spin symmetries. The Dirac bound state energy eigenvalues equation and the corresponding wave functions are found in a closed form. Finally we give some of our conclusions in Section 3.

## 2. The Solution of the Dirac Wave Equation

2.1. *The Spin Symmetry*

The Dirac equation for massive fermions interacting via the scalar $S(r)$ and the time-like vector $V(r)$ potential fields (in $\hbar = c = 1$ units) is [70, 71]:

$$\left[ \vec{\alpha}\cdot\vec{p} + \beta(S(r)+M) - (E_{nm} - V(r)) \right]\Psi(\vec{r}) = 0, \qquad (2)$$

where $\alpha$ and $\beta$ are the usual Dirac matrices, $E_{nm}$ denotes the relativistic energy of the system, $\vec{p} = -i\vec{\nabla}$ denotes the three dimensional momentum operator. Assuming the wave function as $\Psi(\vec{r}) = \begin{pmatrix} f(\vec{r}) \\ g(\vec{r}) \end{pmatrix}$, where $f(\vec{r})$ and $g(\vec{r})$ are the upper and lower spinor components of the Dirac wavefunctions, respectively [72, 73]. Substituting it into Eq. (2), we can obtain two coupled differential equations for the upper and lower radial wave functions. Further, combining the two equations, by applying the spin symmetry for a particle moving in the presence a magnetic and AB flux fields and performing a simple transformation $\vec{p} \to \vec{p} - (e/c)\vec{A}$, we can obtain [60-64, 67]:

$$\left[ \left(\vec{p} - \frac{e}{c}\vec{A}\right)^2 + 2(E_{nm} + M)V(r) - (E_{nm}^2 - M^2) \right] f(\vec{r}) = 0, \qquad (3)$$

where $\vec{A}$ is the vector potential which we take in the form: $\vec{A} = (0, Br/2 + \Phi_{AB}/2\pi r, 0)$ [61, 63, 67]. Now, in choosing $f(\vec{r}) = f_{nm}(r)e^{im\varphi}/\sqrt{r}$ and substituting the potential (1) into Eq. (3), we can get

$$\frac{d^2 f_{nm}(r)}{dr^2} + \left\{ E_{nm}^2 - M^2 + \frac{emB}{2c} - \frac{e^2 B\Phi_{AB}}{4\pi c^2} - \left[ \left(m - \frac{e\Phi_{AB}}{2\pi c}\right)^2 - \frac{1}{4} \right]\frac{1}{r^2} \right.$$
$$\left. + 2(E_{nm} + M)c\frac{1}{r} - 2(E_{nm} + M)br - \left[ 2(E_{nm} + M)a + \frac{e^2 B^2}{4c^2} \right]r^2 \right\} f_{nm}(r) = 0. \qquad (4)$$

Further, by making the change of variable $\chi = \varepsilon_B^{1/4} r$, we can obtain

$$\frac{d^2 f_{nm}(\chi)}{d\chi^2}$$
$$+ \left\{ \frac{\zeta}{\varepsilon_B^{1/2}} - \left(m'^2 - \frac{1}{4}\right)\frac{1}{\chi^2} + \left[\frac{2(E_{nm} + M)c}{\varepsilon_B^{1/4}}\right]\frac{1}{\chi} - \frac{2(E_{nm} + M)b}{\varepsilon_B^{3/4}}\chi - \chi^2 \right\} f_{nm}(\chi) = 0 \qquad (5)$$

where the following assignments $\varepsilon_B = 2a(E_{nm} + M) + \omega_c^2/4$, $\zeta = \left( E_{nm}^2 - M^2 + \frac{m'\omega_c}{2} \right)$, $m' = m - e\Phi_{AB}/2\pi c$ have been used and $\omega_c = eB/c$ is the cyclotron frequency.

It is obvious that Eq. (5) has two singular pointes: the first one is regular at the origin and the second one is irregular at infinity. Therefore, we can write the solutions of the Eq. (5) as expressions valid (authentic) in the neighborhood of both singular points.

Now, with a suitable choice of our ansatz, we can express the upper radial component, $f_{nm}(\chi)$ in the form of $f_{nm}(\chi) = \chi^{m'+1/2} \exp[-\chi(\chi+b')/2] F(\chi)$, and hence Eq. (5), becomes

$$\chi \frac{d^2 F(\chi)}{d\chi^2} + \left[2\left(m' + \frac{1}{2}\right) - b'\chi - 2\chi^2\right] \frac{dF(\chi)}{d\chi}$$
$$+ \left[c' - \left(m' + \frac{1}{2}\right)b' + \left(\frac{\zeta}{\varepsilon_B^{1/2}} + \frac{b'^2}{4} - 2\left(m' + \frac{1}{2}\right) - 1\right)\chi\right] F(\chi) = 0, \quad (6)$$

where the following identifications: $b' = 2(E_{nm} + M)b/\varepsilon_B^{3/4}$ and $c' = 2(E_{nm} + M)c/\varepsilon_B^{1/4}$ were used.

The above differential equation resembles the so-called biconfluent Heun's (BCH) differential equation [74]

$$\frac{d^2 u}{d\zeta^2} + \frac{1}{\zeta}\left[\alpha + 1 - \beta\chi - 2\zeta^2\right]\frac{du}{d\zeta} + \left[\frac{\delta + \beta + \alpha\beta}{2} + (\gamma - \alpha - 2)\zeta\right]\frac{u}{\zeta} = 0, \quad (7)$$

with the well-known Heun's wavefunction solution given by $u = H_B(\alpha, \beta, \gamma, \delta, -\zeta)$. In comparing Eq. (6) with its counterpart Eq. (7), we can conclude that Eq. (6) is simply the BCH differential equation [74], whose solution is BCH function, $H_B$:

$$F(r) = H_B\left(2m', \frac{2(E_{nm}+M)b}{\varepsilon_B^{3/4}}, \frac{\zeta}{\varepsilon_B^{1/2}} + \frac{(E_{nm}+M)^2 b^2}{\varepsilon_B^{3/2}}, \frac{4(E_{nm}+M)c}{\varepsilon_B^{1/4}}, -\varepsilon_B^{1/4} r\right),$$

and consequently the upper radial spinor component of the Dirac wavefunction:

$$f_{nm}(\chi) = \chi^{m'+1/2} \exp\left[-\chi\left(\chi + \frac{2(E_{nm}+M)b}{\varepsilon_B^{3/4}}\right)\bigg/2\right]$$
$$\times H_B\left(2m', \frac{2(E_{nm}+M)b}{\varepsilon_B^{3/4}}, \frac{\zeta}{\varepsilon_B^{1/2}} + \frac{(E_{nm}+M)^2 b^2}{\varepsilon_B^{3/2}}, \frac{4(E_{nm}+M)c}{\varepsilon_B^{1/4}}, -\chi\right). \quad (8)$$

Likewise, the lower radial spinor component of the wavefunction is $g_{nm}(\chi) = \chi^{m'+3/2} \exp[-\chi(\chi + 2(E_{nm}+M)b/\varepsilon_B^{3/4})/2] G(\chi)$, with

$$G(r) = H_B\left(2m'+2, \frac{2(E_{nm}+M)b}{\varepsilon_B^{3/4}}, \frac{\zeta}{\varepsilon_B^{1/2}} + \frac{(E_{nm}+M)^2 b^2}{\varepsilon_B^{3/2}}, \frac{4(E_{nm}+M)c}{\varepsilon_B^{1/4}}, -\varepsilon_B^{1/4} r\right). \quad (9)$$

Let us now follow the results given in Eq. (6) by defining the parameters $P, Q, R$, and the function $\tilde{F}$ representing $F$ as follows:

$$P = m' + \frac{1}{2}, \qquad Q = 1 - \left(m' + \frac{1}{2}\right)\frac{b'}{c'}, \qquad \text{if} \qquad \tilde{F} = F,$$

$$R = \frac{\zeta}{\varepsilon_B^{1/2}} + \frac{b'^2}{4} - 2\left(m' + \frac{1}{2}\right) - 1, \qquad \text{if} \qquad \tilde{F} = F, \tag{10}$$

with the above definitions, Eq. (6) can be simplified as

$$\chi \frac{d^2 \tilde{F}(\chi)}{d\chi^2} + [2P - b'\chi - 2\chi^2]\frac{d\tilde{F}(\chi)}{d\chi} + (R\chi - Qc')\tilde{F}(\chi) = 0. \tag{11}$$

In order to solve the above differential equation, the function $\tilde{F}(\chi)$ is assumed to be in the Frobenius series form as $\tilde{F}(\chi) = \sum_n C_n \chi^n$. So we can plug it into Eq. (11) and hence obtain the recurrence relation:

$$C_{n+2} = \frac{[Qc' - b'(n+1)]C_{n+1} - (R - 2n)C_n}{(n+2)(n+2P+1)}. \tag{12}$$

Assuming $C_{-1} = 0$ and $C_0 = 1$, then the first three coefficients of the recurrence relation (12) can be obtained as follows:

$$C_1 = \frac{Qc'}{2P}, \qquad C_2 = \frac{(Qc' - b')\frac{Qc'}{2P} - R}{2(2P+1)},$$

$$C_3 = \frac{1}{6(P+1)}\left\{\frac{Qc' - 2b'}{2(2P+1)}\left[\frac{Qc'}{2P}(Qc' - b') - R\right] - \frac{Qc'}{2P}(R - 2)\right\}. \tag{13}$$

At this stage, we can obtain the analytical solution to the radial part of Dirac equation. This can be accomplished by breaking the series (12) of the BCH function into Heun's polynomial of degree $n$. Imposing the following conditions on the two coefficients: $C_{n+1} = 0$ and $R = \gamma - \alpha - 2 = 2n$ with $n = 1, 2, 3, \ldots$. So from the condition $R = 2n$, it is possible to obtain a formal expression for the energy eigenvalues. Therefore, after adopting the aforementioned limitation, we can simply obtain the required energy eigenvalues. Also, the necessary condition the BCH series,

Eq. (7), becomes a polynomial of degree $n$ with $\gamma = 2n + \alpha + 2$, [75, 76]

$$\sqrt{1 - 4\left(\frac{1}{4} - m^2 - \frac{e^2\Phi_{AB}^2}{4\pi^2 c^2} + \frac{e\Phi_{AB}}{2\pi c}\right)} + 2n + 2$$

$$- \frac{E_{nm}^2 - M^2 + \frac{emB}{2c} - \frac{e^2 B\Phi_{AB}}{2\pi c^2}}{\sqrt{2a(E+M) + \frac{e^2 B^2}{4c^2}}} - \frac{(E_{nm} + M)^2 b^2}{\left[2a(E+M) + \frac{e^2 B^2}{4c^2}\right]\sqrt{2a(E+M) + \frac{e^2 B^2}{4c^2}}} = 0,$$

or

$$-2\sqrt{2a(E+M) + \frac{e^2 B^2}{4c^2}}\left[\frac{1 + \sqrt{1 - 4\left(\frac{1}{4} - m^2 - \frac{e^2\Phi_{AB}^2}{4\pi^2 c^2} + \frac{e\Phi_{AB}}{2\pi c}\right)}}{2} + n + \frac{1}{2}\right]$$

$$+ E_{nm}^2 - M^2 + \frac{emB}{2c} - \frac{e^2 B\Phi_{AB}}{2\pi c^2} + \frac{(E_{nm} + M)^2 b^2}{2a(E+M) + \frac{e^2 B^2}{4c^2}} = 0. \quad (14)$$

where $n = 1, 2, 3, ...$ is the radial quantum number. For the sake of completeness in this discussion and making one is confident on the variety of results. It is found that Eq. (9) of Ref. [62] is same as the present one in Eq. (4), therefore we can impose same condition in Eq. (18) of Ref. [62].

Alternatively, making the following substitutions: $m' \to \beta$, $\zeta \to -\varepsilon^2$, $\varepsilon_B \to (p^2 = \gamma^2)$, $2(E_{nm} + M)b \to 2pq$ and $s \to n$, it can easily provide Eq. (14). That is, the above equation is correct in its present form.

In table 1, we have obtained the energies of the $n=1,2,3$ and 4 states in the spin symmetric case. To see the effect of the potential parameters on these states, we have fixed the value of the parameter $b$ while have increased the value of the parameter which could lead to an increasing in the energy states. Further, when we have increased $b$, the energy levels are found to be slightly decreasing. The energy shift between the different successive $n>1$ states is found evenly spaced. The increase in the strength of this magnetic field would lead to a wider shift. The excited energy states go up wider shift nearly 0.11 *MeV* to 0.14 *MeV* when the strength of the magnetic field $B$ changes from 1.0 *T* to 1.5 *T*.

To show the effect of changing the Killingbeck model parameters on the energy states, we have plotted the energy versus the potential parameters $a$ and $b$ for various values of the magnetic field strength $B$ as shown in Figures 1 and 2, respectively.

## 2. 2. *The Pseudo-Spin Symmetry*

In this section, we begin by studying the pseudospin symmetric case. So we need to solve the following Dirac equation

$$\left[\left(\vec{p}-\frac{e}{c}\vec{A}\right)^2 + 2(E_{nm}-M)V(r)-\left(E_{nm}^2-M^2\right)\right]g(\vec{r}) = 0. \tag{15}$$

So to avoid any repetition in our solution to Eq. (14) and to obtain a similar solution as before in subsection 2.1, the pseudo-spin symmetry can be found by making the following transformations $f(r) \to g(r)$, $E_{nm} \to -E_{nm}$ and $V(r) \to -V(r)$ into Eq. (14) and obtain

$$2n+2+2m-\frac{e\Phi_{AB}}{\pi c}+\frac{E_{nm}^2-M^2+\frac{emB}{2c}-\frac{e^2 B\Phi_{AB}}{2\pi c^2}}{\sqrt{2a(E_{nm}-M)+\frac{e^2 B^2}{4c^2}}}$$
$$-\frac{(E_{nm}-M)^2 b^2}{\sqrt{2a(E_{nm}-M)+\frac{e^2 B^2}{4c^2}}\left[2a(E_{nm}-M)+\frac{e^2 B^2}{4c^2}\right]} = 0, \tag{16}$$

and hence the wave function for the pseudo-spin symmetry is taken as:

$$g_{nm}(\chi) = \chi^{m'+1/2} \exp\left[-\chi\left(\chi+\frac{2(E_{nm}-M)}{\varepsilon_B^{3/4}}\right)\bigg/2\right]$$
$$\times G_B\left(2m', \frac{2(E_{nm}-M)b}{\varepsilon_B^{3/4}}, \frac{\zeta}{\varepsilon_B^{1/2}}+\frac{2(E_{nm}-M)^2 b^2}{\varepsilon_B^{3/2}}, \frac{2(E_{nm}-M)b}{\varepsilon_B^{3/4}}, -\chi\right). \tag{17}$$

In table 2, we have also obtained the energy states of $n=1,2,3$ and 4, in the pseudo-spin symmetric case, for various values of the potential parameters. We fixed the value of the parameter $b$ while changed the value of the parameter a, it is obvious that the energy increases by increasing a. Also increasing $b$ leads to slightly decreasing energy. This behavior is very similar to spin symmetric case. So by applying a magnetic field this leads to an increasing in the excited states shift with the increasing $B$. It is noted that the shift, under the effect of $B$, is slightly smaller than the shift in the spin symmetric case.

For example, the excited energy states go up shift nearly of 0.09 *MeV* to 0.13 *MeV* when the strength of the magnetic field $B$ changes from 1.0 *T* to 1.5 *T*.

We have also plotted the energy versus the potential parameters *a* and *b* for various values of the magnetic field strength *B* as shown in Figures 3 and 4, respectively. The pseudo-spin symmetry is seen a very similar to that in spin symmetry, Figures 1 and 2.

In Figure 5, we plot the ground state energy versus the magnetic field $B$ for three different values of the Aharonov-Bohm flux field. It is obvious that the energy increases with the increasing magnetic field in the spin symmetric case. In Figure 6, we also plot the ground state energy versus the magnetic field $B$ for three different values of the Aharonov-Bohm flux field. It is obvious that the energy increases linearly with the increasing magnetic field in the pseudo-spin symmetric case.

In Figure 7. we plot the spin symmetric energy versus the magnetic field $B$ for three different energy states. The energy increases with increasing magnetic field for all states. Further. in Figure 8, we also plot the pseudo-spin symmetric energy versus the magnetic field $B$ for three different energy states. The energy increases linearly with the increasing of the magnetic field for all states.

## 3. Discussions and Results

In this work, we have obtained exact analytical solutions of the Dirac equation for the Killingbeck potential in the presence of the external magnetic and Aharonov-Bohm (*AB*) flux fields under the spin and pseudo-spin symmetries.

The only role of a magnetic field consists in reducing the angular frequency and the entire dynamics of the system remains unchanged.

We have found that the energy is increasing with the increasing of magnetic field strength. The energy also changes with increasing the magnetic flux density. The spin symmetric solution is very similar to the pseudo-spin symmetric one with a slight difference.

Let us comment on our results in Figures 1 and 2 that when taking fixed values of *a* and *b*, the energy increases with the increasing magnetic field strength.

Further, in Figures 3 and 4, notice that for fixed values of *a* and *b*, the energy increases when the magnetic field grows.

Also, in Figures 5 and 6, notice that for a fixed value of the magnetic field, the energy increases when the magnetic flux density grows. Finally, in Figure 7 and 8, notice that for a fixed value of the magnetic field, the energy increases when the quantum number n grows.

**References**

[1] Tan Z.L., Park C.-H. and Louie S.G. *Phys. Rev. B* **81** (2010) 195426.

[2] Yannouleas C., Romanovsky I. and Landman U. *Phys. Rev. B* **89** (2014) 035432.

[3] Salamin Y.I., Hu S., Hatsagortsyan K.Z. and Keitel C.H. *Phys. Rep.* **427**(2-3) (2006) 41.

[4] Szpak N. and Schutzhold R. *New J. Phys.* **14** (2012) 035001.

[5] Lamata L., Casanova J., Gerritsma R., Roos C.F., Garcia-Ripoll J.J. and Solano E. *New J. Phys.* **13** (2011) 1367.

[6] Witthaut D., Salger T., Kling S., Crossert C. and Weitz M. *Phys. Rev. A* **84** (2011) 033601.

[7] Qi X.L. and Zhang S.C. *Rev. Mod. Phys.* **83** (2011) 1057.

[8] Bernevig B.A., Hughes T.L. and Zhang S.C. *Science* **314** (2006) 1757.

[9] Arima A., Harvey M. and Shimizu K. *Phys. Lett. B* **30** (1969) 517.

[10] Hect K.T. and Adler A. *Nucl. Phys. A* **137** (1969) 129.

[11] Bell J.S. and Ruegg H. *Nucl. Phys. B* **98** (1975) 151. ; Errata *Nucl. Phys. B* 104 (1976) 546.

[12] Ginocchio J.N. *Phys. Rev. Lett.* **78** (1997) 436.

[13] Ginocchio J.N. Leviatan A., Meng J. and Zhou S.G. *Phys. Rev. C* **69** (2004) 034303.

[14] Leviatan A. and Ginocchio J.N. *Phys. Lett. B* **518** (2001) 2140.

[15] Bahri C., Draayer J.P. and Moszkowski S.A. *Phys. Rev Lett.* **68** (1992) 2133.

[16] Blokhin A.L. Bahri C. and Draayer J.P. *Phys. Rev. Lett.* **74** (1995) 4149.

[17] Blokhin A.L., Bahri C. Draayer J.P. *J. Phys. A* **29** (1996) 2039.

[18] Ginocchio J.N. *Phys. Rev. C* **69** (2004) 034318.

[19] Hexel O., Jensen H.D. and Suees H.E. *Phys. Rev. E* **75** (1949) 1766.

[20] Meng J., Sugawara-Tanabe K., Yamaji S., Ring P. and Arima A. *Phys. Rev C* **59** (1999) 154.

[21] Troltenier D., Bahri C. and Draayer J.P. *Nucl. Phys. A* **586** (1995) 53.

[22] Ikhdair S.M. and Sever R. *Appl. Math. Comput.* **216** (2010) 911.

[23] Berkdemir C. *Nucl. Phys. A* **770** (2006) 32.

[24] Dong S.-H. and Wei G.-F. *Europhys. Lett.* **87** (2009) 40004.

[25] Dong S.-H. and Wei G.-F. *Phys. Lett. B* **686** (2010) 288.

[26] Dong S.-H. and Wei G.-F. *Eur. Phys. J. A* **46** (2010) 207.

[27] Bohr A., Hamamoto I. and Mottelson B.R. *Phys. Scr.* **26** (1982) 267.

[28] Dudek J., Nazarewicz W., Szymanski Z. and Leander G.A. *Phys. Rev. Lett.* **59** (1987) 1405.

[29] Stuchbery A.E. *Nucl. Phys. A* **700** (2002) 83.

[30] Stephens et al, Phys. *Rev. Lett.* **65** (1990) 301.

[31] Liang H., Zhao P., Zhang Y., Meng J. and Giai V.V. *Phys. Rev C* **83** (2011) 041301(R).

[32] Liang H., Shen S., hao P. and Meng J. *Phys. Rev C* **87** (2013) 014334.

[33] Lu B.-N., Zhao E.-G. and Zhou S.-G. *Phys. Rev C* **88** (2013) 024323.

[34] Alhaidari A.D., Bahlouli H. and Al-Hasan A. *Phys. Lett. A* **349** (2006) 87.

[35] Meng J., Sugawara-Tanabek K., Yamaji S., Ring P. and Arima A. *Phys. Rev C* **58** (1998) R628.

[36] Ginocchio J.N. and Leviatan A. *Phys. Lett. B* **425** (1998) 1.

[37] Zhou S.G., Meng J. and Ring P. *Phys. Rev. Lett*. **91** (2003) 262501.

[38] Liang H., Long W.H., Meng J. Giai V. Eur. Phys. J. A **44** (2010) 119.

[39] Meng J., Sugawara-Tanabe K., Yamaij S. and Arima A. *Phys. Rev. C* **59** (1999) 154.

[40] Shen S., Liang H., Zhao P., Zhang S. and Meng J. *Phys. Rev. C* **88** (2013) 024311.

[41] Alberto P., Malheiro M., Frederico T. and de Castro A.S. *Phys. Rev. A* **92** (2015) 062137.

[42] Lisboa R., Malheiro M., Alberto P., Fiolhais M. and de Castro A.S. *Phys. Rev. C* **81** (2010) 064324.

[43] Marcos S., Lopez-Quelle M., Niembro R. and Savushkin L.N. *Eur. Phys. J. A* **37** (2008) 251.

[44] Castro L.B. and de Castro A.S. *Chin. Phys. B* **23** (2014) 090301.

[45] Alhaidari A.D. *Phys. Lett. B* **699** (2011) 309.

[46] Castro L.B., de Castro A.S. and Alberto P. *Ann. Phys.* (NY) **356** (2015) 83.

[47] Li F.-Q., Zhao P.-W. and Liang H.-Z. *Chin. Phys. C* **35** (2011) 825.

[48] Zhao X.-Q., Jia C.-S. yang Q.-B. *Phys. Lett. A* **337** (2005) 189.

[49] Agboola D. *Few-Body Syst*. **52** (2012) 31.

[50] Zhang M.-C. and Huang-Fu G.-Q. *Ann. Phys*. **327** (2012) 841.

[51] Tokmehdashi H., Rajabi A.A. and Hamzavi M. *J. Theor. Appl. Phys.* **9** (2015) 15.

[52] Akcay H. *Phys. Lett. A* **373** (2009) 616.

[53] Dong S.-H. and Lozada-cassou M. *Phys. Lett. A* **330** (2004) 168.

[54] Arda A., Tezcan C. and Sever R. *Few-Body Syst*. **57** (2016) 93.

[55] Eshghi M. and Ikhdair S.M. *Math. Meth. Appl. Sci*. **37** (2014) 2829.

[56] Eshghi M. and Ikhdair S.M. *Chin. Phys. B* **23**(12) (2014) 120304.

[57] Eshghi M. *Can. J. Phys.* **91** (2013) 71–74.

[58] Eshghi M. and Mehraban H. *Chin. J. Phys.* **50**(4) (2012) 533.

[59] Hamzavi M. and Rajabi A.A. *Can. J. Phys*. **91** (2013) 411.

[60] Ikhdair S.M. and Falaye B.J. *J. Assoc. Arab Uni. Bas. Appl. Sci.* **16** (2014) 1.

[61] Ikhdair S.M. and Sever R. *Adv. High Energy Phys*. **2013** (2013) 562959.

[62] Ikhdair S.M., Babatunde J.F. and Hamzavi M. *Ann. Phys.* **353** (2015) 282.

[63] Ikhdair S.M. *Adv. High Energy Phys*. **2013** (2013) 491648.

[64] Hamzavi M., Ikhdair S.M. and Babatunde J.F. *Ann. Phys.* **341** (2014) 153.

[65] Ikhdair S.M. *Eur. Phys. J. A* **40** (2009) 143.

[66] Ikhdair S.M. *Eur. Phys. J. A* **39** (2009) 307.

[67] Sharifi Z., Tajic F., Hamzavi M. and Ikhdair S.M. *Z. Naturforsch.* **70**(7)a (2015) 499.

[68] Neyazi H., Rajabi A.A. and Hassanabadi H. *Nucl. Phys. A* (2015), http://dx.doi.org/10.1016/j.nuclphysa.2015.08.007

[69] Killingbeck J. *Phys. Lett. A* **65** (1987) 87.

[70] Ginocchio J.N. *Phys. Rep.* **414** (2005) 165.

[71] Ginocchio J.N. and Leviatan A., *Phys. Lett. B* **425** (1998) 1.

[72] Falaye B.J., Dong S-H., Oyewumi K.J., Ilaiwi K.F, and Ikhdair S.M., *Int. J. Mod. Phys. E* **24** (11) (2015) 1550087.

[73] Chang-Yuan C., *Phys. Lett. A* **339** (2005) 283.

[74] Slavyanov S. Yu. and Lay W., "*Special Functions: A Unifield Theory Based in Singularities*", Oxford University Press, New York, 2000.

[75] Ronveaux A., "*Heun's Differntial Equations*", Oxford University Press, Oxford, 1995.

[76] Gurappa N. and Panigrahi P.K. *J. Phys. A* **37** (2004) L605.

Table 1. The spin-symmetric energy states of the Dirac particle in the Killingbeck potential field.

| n | b | a | $E_s$ (MeV) ($\Phi = 2.0$ T, $M = 5.0$ MeV, $m = e = c = 1$) | | |
|---|---|---|---|---|---|
| | | | $B = 1.0$ T | $B = 1.2$ T | $B = 1.5$ T |
| n = 1 | 0.005 | 0.001 | 5.273485251 | 5.323036087 | 5.397166864 |
| | | 0.003 | 5.294470611 | 5.340782661 | 5.411507210 |
| | | 0.005 | 5.314084649 | 5.357668430 | 5.425374208 |
| | 0.007 | 0.001 | 5.272596822 | 5.322404523 | 5.396754568 |
| | | 0.003 | 5.293698069 | 5.340212048 | 5.411121852 |
| | | 0.005 | 5.313401395 | 5.357148110 | 5.425012512 |
| | 0.009 | 0.001 | 5.271412478 | 5.321562555 | 5.396204891 |
| | | 0.003 | 5.292668162 | 5.339451317 | 5.410608083 |
| | | 0.005 | 5.312490493 | 5.356454415 | 5.424530285 |
| n = 2 | 0.009 | 0.005 | 5.423823650 | 5.482709813 | 5.573630596 |
| n = 3 | 0.009 | 0.005 | 5.533240322 | 5.606436282 | 5.719144859 |
| n = 4 | 0.009 | 0.005 | 5.640843664 | 5.727788566 | 5.861330050 |

**Table 2.** The pseudo-spin symmetric energy states of a Dirac particle in Killingbeck potential field.

| | | | $E_{Ps}$ (MeV) ($\Phi = 2.0\ T$, $M = 5.0\ MeV$, $m = e = c = 1$) | | |
|---|---|---|---|---|---|
| | $b$ | $a$ | $B = 1.0\ T$ | $B = 1.2\ T$ | $B = 1.5\ T$ |
| $n = 1$ | 0.005 | 0.001 | 5.263670129 | 5.314776171 | 5.390526897 |
| | | 0.003 | 5.264248822 | 5.315346229 | 5.391084622 |
| | | 0.005 | 5.264828765 | 5.315917291 | 5.391643106 |
| | 0.007 | 0.001 | 5.263669496 | 5.314775550 | 5.390526294 |
| | | 0.003 | 5.264248187 | 5.315345607 | 5.391084018 |
| | | 0.005 | 5.264828129 | 5.315916668 | 5.391642502 |
| | 0.009 | 0.001 | 5.263668651 | 5.314774722 | 5.390525490 |
| | | 0.003 | 5.264247341 | 5.315344778 | 5.391083213 |
| | | 0.005 | 5.264827281 | 5.315915837 | 5.391641695 |
| $n = 2$ | 0.009 | 0.005 | 5.360133315 | 5.428762792 | 5.530111950 |
| $n = 3$ | 0.009 | 0.005 | 5.454107542 | 5.539632933 | 5.665504273 |
| $n = 4$ | 0.009 | 0.005 | 5.546810297 | 5.648634249 | 5.798024647 |

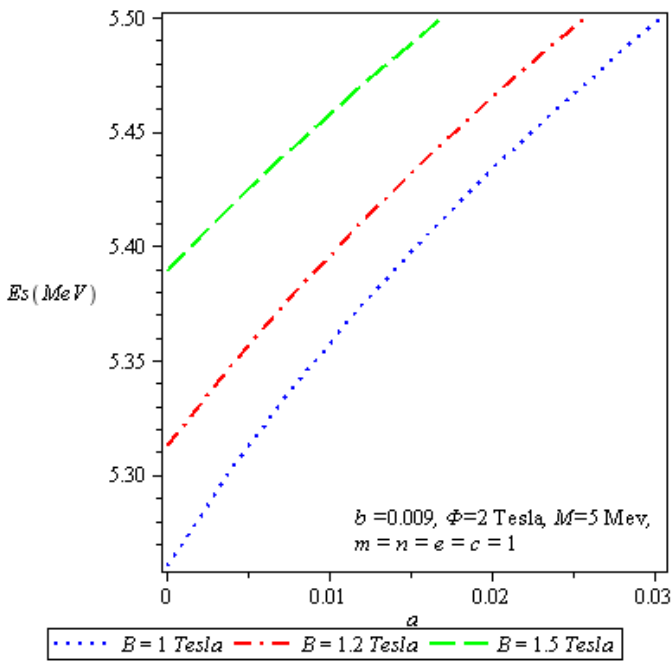

**Fig. 1.** ............

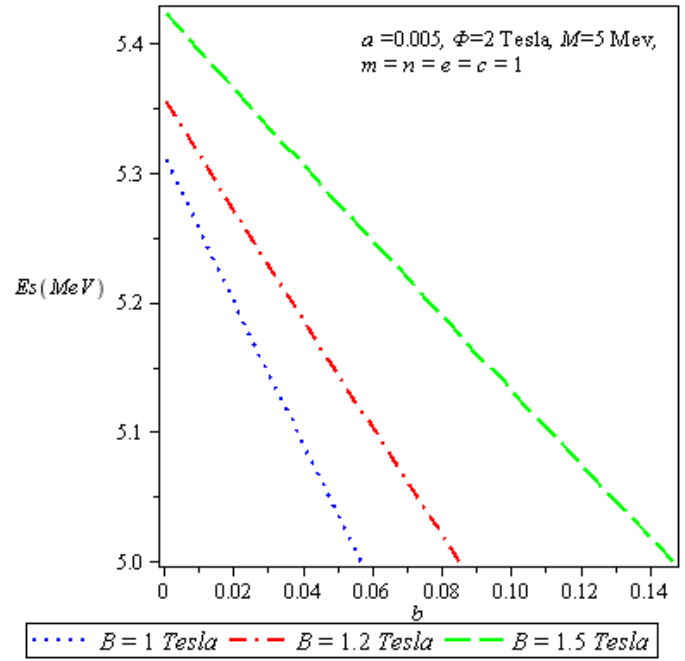

**Fig. 2.** ............

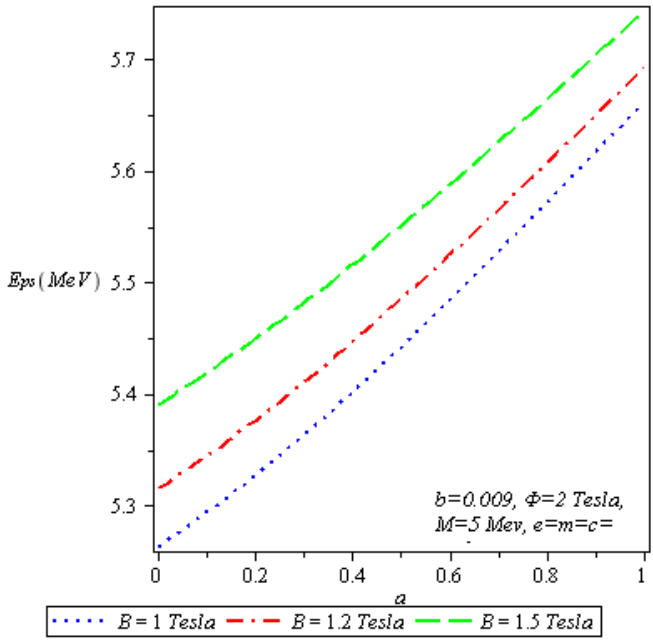

Fig. 3. ............

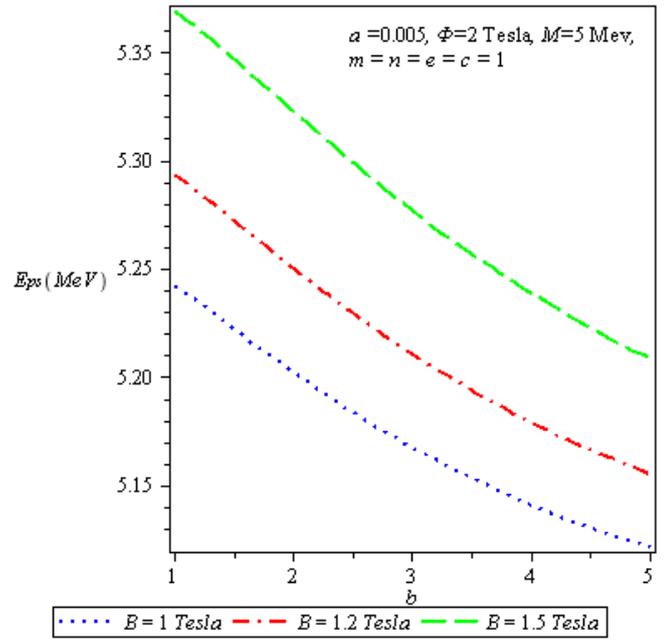

Fig. 4. ............

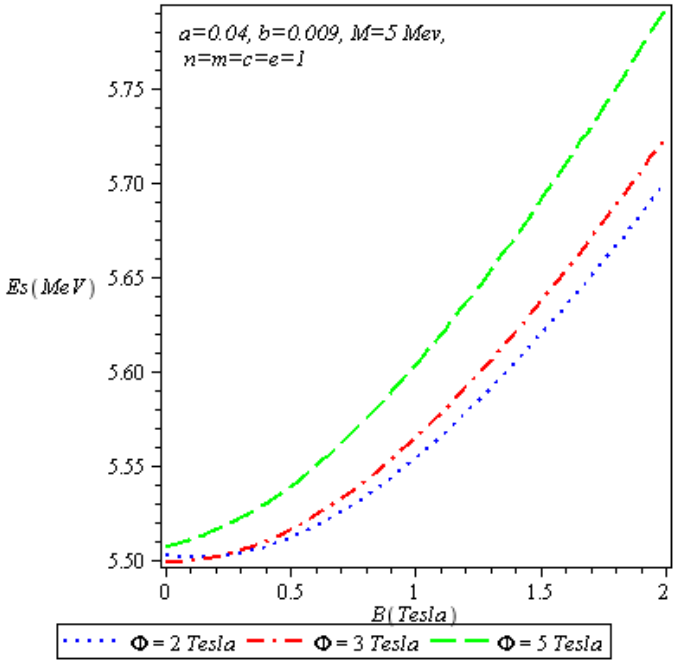

Fig. 5. ............

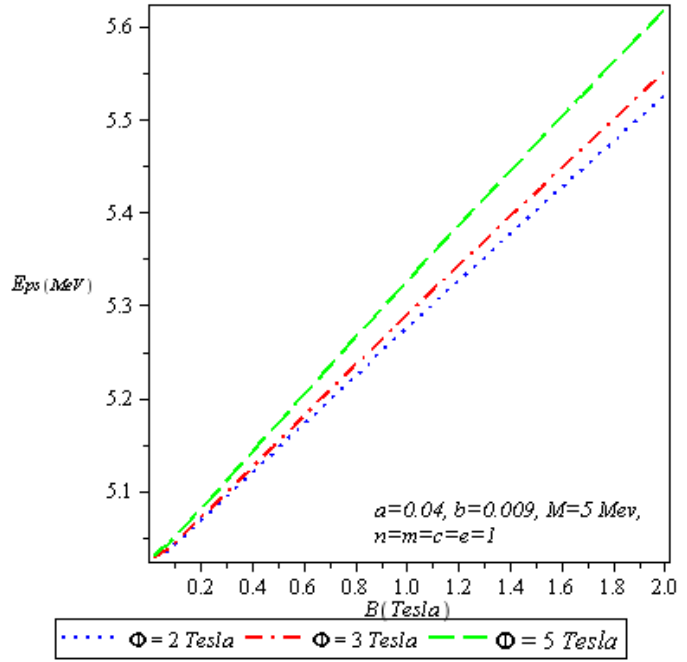

Fig. 6. ............

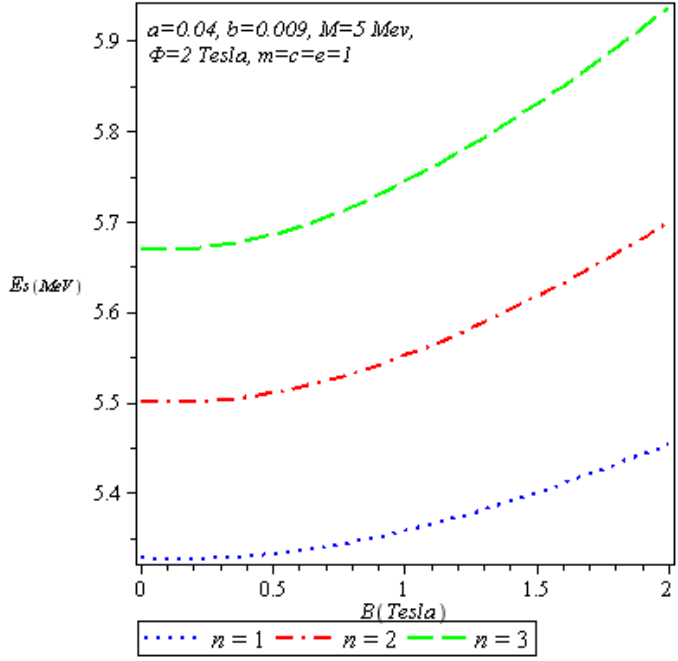 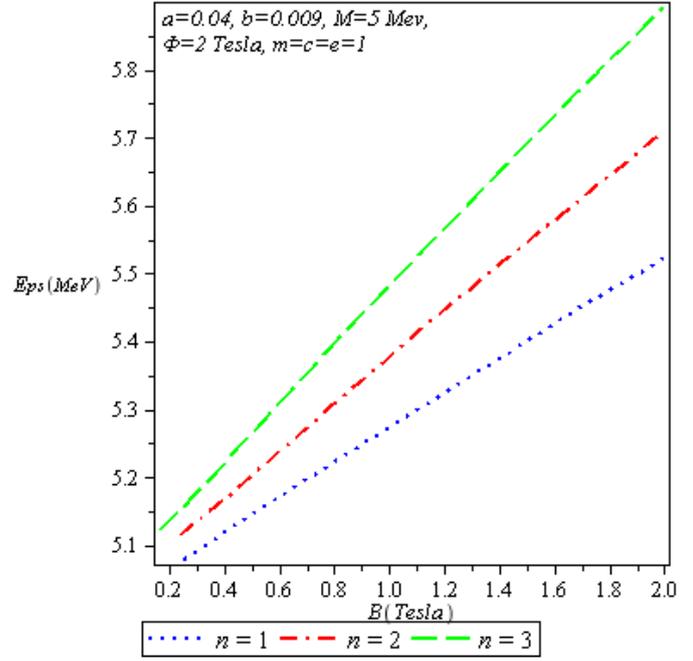

**Fig. 7.** ............  **Fig. 8.** ............